# Social and natural sciences differ in their research strategies, adapted to work for different knowledge landscapes

Klaus Jaffe *
Departamento de Biología de Organismos, Universidad Simón Bolívar, Caracas, Estado Miranda, Venezuela.
email: kjaffe@usb.ve

**SUMMARY**

Do different fields of knowledge require different research strategies? A numerical model exploring different virtual knowledge landscapes, revealed two diverging optimal search strategies. Trend following is maximized when the popularity of new discoveries determine the number of individuals researching it. This strategy works best when many researchers explore few large areas of knowledge. In contrast, individuals or small groups of researchers are better in discovering small bits of information in dispersed knowledge landscapes. Bibliometric data of scientific publications showed a continuous bipolar distribution of these strategies, ranging from natural sciences, with highly cited publications in journals containing a large number of articles, to the social sciences, with rarely cited publications in many journals containing a small number of articles. The natural sciences seem to adapt their research strategies to landscapes with large concentrated knowledge clusters, whereas social sciences seem to have adapted to search in landscapes with many small isolated knowledge clusters. Similar bipolar distributions were obtained when comparing levels of insularity estimated by indicators of international collaboration and levels of country-self citations: researchers in academic areas with many journals such as social sciences, arts and humanities, were the most isolated, and that was true in different regions of the world. The work shows that quantitative measures estimating differences between academic disciplines improve our understanding of different research strategies, eventually helping interdisciplinary research and may be also help improve science policies worldwide.

**INTRODUCTION**

The maturation of empirical science, as catalyzed by Galileo Galilei, was fundamental in triggering the industrial revolution, the most significant transformation of human society in the last 10000 years. A large effort is now invested in analyzing scientific productivity and its social dynamics. Entire journals are dedicated to it, such as Scientometric and the Journal of Informetrics. The dynamics of research in the natural and social sciences however diverges. This has been recognized long ago [1, for example] but pinpointing precise quantitative differences between both sciences has been an elusive endeavor. Citations are the most common tool nowadays to estimate the scientific quality of a researcher or paper, but this method that has had important limitations since its beginnings [2, for example]. Few efforts analyze the mechanics through which the scientific method works best. The scientific method relies on several key elements for its smooth working [3]. Among them are: 1- Humility to recognize that our mind is very limited in grasping the underlying dynamics of natural phenomena and needs the experiment or an empirical feedback to keep if from digressing into the absurd; 2- Science is not interested in absolute truth but in incremental advancements in our understanding of nature; 3- Mathematics as the best language to ask nature our questions.

How do we decide to use scarce financial resources to finance many different small projects, or to concentrate resources in a few "strategically important" projects? How do we know if criteria we use to evaluate and finance projects in the natural sciences also work adequately in the social sciences? We actually have no clear answer to these questions at the moment. Different sciences and scientific disciplines cultivate different values and attitudes and show differences in quantifiable characteristics [4-7]. We also know that the development of different scientific disciplines has different effects on economic growth. For example, the subject areas with the largest relative number of publication in wealthy countries today are neuroscience and psychology; investment in these areas however does not produce economic growth in less developed countries. In contrast, middle income countries that give more value to basic natural science in a given time period show faster economic growth in the following years [8-9]. Additionally, countries whose researchers are less provincial and cite more works from countries different to theirs (have fewer country self-citations) are also those whose scientists produce relatively lower numbers of author self-citations. These countries are the ones producing scientific papers with higher overall citation impact [10].

To understand the underlying dynamics of this phenomena I propose to use the concept of knowledge landscapes. This concept has been developed in explaining decision making mechanisms in foraging strategies in ants [11], in artificial intelligence, swarm intelligence, and many other settings [12], suggesting that optimal foraging strategies address fundamental aspects of complex dynamics present in a large variety of situations. Here, the results of agent based computer simulations build to analyze the efficiency of different foraging strategies for exploring various landscapes [11], were generalized and adapted to understand aspects of different publication strategies in scientific research. The results could be validated with empirical observations, opening a novel way to understand differences among scientific disciplines that are relevant for future science policy.

## METHODS

**The Model**

The numerical model originally coded in Fortran, was developed for studying optimal foraging strategies in ants in different resource landscapes. The results of the model were published long ago [11, 13] and were validated experimentally by different researchers [see review in 14]. A game written in Java-script, based on this model, can be accessed on the web (http://atta.labb.usb.ve/SmartAnts.html). The simulation consisted in foragers exploring different landscapes using random walks. Once a resource was discovered, they returned to the nest recruiting nestmates so as to collect the resources discovered. Two possible decision making systems used for recruitment were tested: The Democratic system or trend following, were all workers eventually perform all tasks, and were the first discovery will draw the most recrutees; and the Technocratic system, where workers specialize either in scouting or in retrieval and were the society collects several smaller resources simultaneously. Here specialized workers signal the palatability, quality, or quantity of a resource regulating the amount of recrutees for each source using simple decision rules. The results of this model [11] were generalized to study strategies exploring different knowledge landscapes. Knowledge landscapes can have different forms and have also been called "knowscapes" [12]. The study of optimal foraging and recruitment strategies of ants have helped in the past in developing heuristic programs such as "swarm intelligence" [15] "active walks" [16-17], and adaptive landscapes, used to study dynamic complex systems. The approach attempted here is an extension of these efforts.

The generalization of the knowledge landscape as applied to academic research is as follows: We assume a scientific community displaying central-place foraging in a finite area with 200 researchers. In each simulation run, a fixed amount of resources were randomly distributed in the landscape. Each researcher explored independently the knowledge landscape, and was randomly assigned as a original researcher or "leader" or a less audacious researcher or "follower", according to a fixed proportion (Ro). Each researcher was randomly assigned two thresholds which defined the decision making system for recruitment they were going to use: one for responding to high impact publications, and another which regulated the initiation of knowledge explorations according to the "quality" of the newly discovered knowledge and the number of researchers engaged in exploring it. Leaders were then left to roam the space and to recruit followers. The speed with which the total amount of resources were discovered was computed. Different patterns of resource distribution in the landscape were tested with different decision making mechanisms for recruiting.

A two-dimensional concentric space with 500 possible sites was modeled. The community of scientists was located at the center of the space. Each researcher was able to explore randomly in the knowledge space, one space at each time interval. Each knowledge cluster was located at one single site. Any number of researchers could share a single space. This simplification stress the fact that, relative to the size of the knowledge horizon of the researchers, the knowledge landscape is enormous, and that the space occupied by knowledge clusters is negligible compared to the size of the knowledge landscape. Knowledge was randomly distributed in the environment. Different densities of knowledge clusters (D) were tested; the cluster size (CS) at each locality could vary.

*Numerical variables used*
N = Number of researchers in the community
Ro = Percent of original researcher in a community: 100-Ro = Rt =Trend Followers (%)
ORo = Optimum percentage of original researchers, i.e. percentage of Ro that allowed the discovery of the maximum amount of knowledge during a fixed time period.
D = Cluster density: mean knowledge cluster density (sites occupied with knowledge clusters (%))
K = Absolute amount of knowledge discovered during a given period
Kc = Knowledge cluster size: (in equivalent of number of trend following researchers required to explore it)
GS = Group Size for Group exploration

*Searching behavior and recruitment*
Researchers were programmed to have a sequence of five behaviors:
1: Random searching for knowledge
2: After finding knowledge, publishing the findings as fast as possible
3: Rt were recruiting depending on the recruitment technique simulated
4: More knowledge was discovered and the published, recruiting even more researchers

*Three different recruitment techniques were modeled:*
1. Group recruitment (GR): Ro always recruited a fixed number of Rt
2. Technocratic recruitment (TR): Ro recruited the exact number of Rt required to research the new knowledge.

3. Democratic recruitment (DR): Ro recruited Rt by promoting their findings. The more Ro the more Rt recruited

**Bibliometry**

Publicly available quantitative scientometric variables were computed from 21135 journals, for 20 different subject areas, grouped by Scopus, and reported by SCImago [18]. The data extracted from SCimago was pooled and is available as data in S1. Different statistical analysis revealed similar trends (see 8). Here we present only graphs with data whose extremes are highly statistically significantly different using non parametric analysis. The time period chosen to sample the data was the year 2011, to guarantees a more or less uniform bibliometric methodology and enough time for the data pools to have retrieved most of the corresponding data (some journal issues appear years after their listed publication year). Total Journals per subject and Total Document were obtained by totalizing data of the respective column. The total Cites for the last 3 years was divided by the total Citable Documents for the last 3 years to obtain Cites / Document. Other variables analyzed are summarized in Table 1

## RESULTS

**Simulations**

Figure 1 shows a typical example of the relation between the amount of knowledge K retrieved by a scientific community during a fixed period for a given amount of researchers. We see that with small knowledge clusters (Kc=5), simulating landscapes with many small knowledge clusters, the curves tended to be broader and the total amount of knowledge retrieved (K) lower compared to landscapes with a few large knowledge clusters (Kc=125). In this last case, the total amount of knowledge discovered was much larger then in situations with many small knowledge clusters, for all proportion of leaders (Ro) simulated.

Figure 2 shows that the optimal number of original researchers in the community decreases with increasing total number of researchers. That is, original research has a larger effect on the amount of knowledge retrieved in small scientific communities than in larger ones.

Figures 3 and 4 show the effect on knowledge retrieval (K) of different recruitment techniques. For large knowledge clusters, TR and DR retrieve the largest amounts of knowledge. Information dispersed in small knowledge clusters are better explored using GR and TR. This shows that the best recruitment technique to be used depends on the nature of knowledge to be explored.: TR produces the best outcome in any situation; DR is useful if knowledge is concentrated in large clusters; and GR works nicely in dispersed knowledge landscapes.

In summary, the results show that depending on the form of distribution of knowledge in the landscape (size and number of knowledge clusters), different recruitment techniques are optimal. Few large knowledge clusters are best discovered with the Democratic decision system, where individuals discovering new knowledge will recruit followers at their maximal capacity. In contrast, the Technocratic system works best when knowledge is distributed in many small knowledge clusters. Here leaders adjust their recruitment effort according to the size of cluster discovered.

**Empirical bibliographic evidence**

The two strategies described above have a mirror in academic disciplines. Certain disciplines focus on a few general basic problems that are the same everywhere, whereas other disciplines have many sub-disciplines, each focusing on a specific problem that might vary locally. These strategies, if they exist in science, should show different publication patterns.

The bibliometric data of scientific publications in different fields presented in Figure 5 shows a continuous gradient between these two type of dynamics. The two extremes of the gradient are the social sciences using journals with and average of less than 60 articles per journal; and the natural sciences with journal publishing over 80 articles per journal and up to 500 articles per journal. Social sciences and arts and humanities had less than one citation per document, whereas multidisciplinary sciences, neuroscience and chemistry, all subject areas from the natural sciences, had more than 3 citations per document in average (See Figure 6). The two extreme cases are Mutidisiplinary science, with high citation rates from colleagues publishing in a few journals containing a large number of articles; and Social Sciences publications with few citations from colleagues publishing in many different journals containing each a low number of articles. Disciplines such as Physics, Chemistry and Material sciences resemble more the pattern of Mutidisiplinary science regarding the use of journals publishing many articles, than that of the Social Sciences. The Arts and Humanities, Economics, Psychology and Business, on the other hand, use journals with few articles each, resembling the pattern of the Social Sciences. Applied sciences and Mathematics have an intermediate ranking regarding the number of documents per journal and number of citations per article. Very closely defined areas such as Dentistry and Veterinary sciences, for example, report fewer journals than broadly defined disciplines such as Medicine which had the largest number of journal reported by Scopus in 2011. If all sub-areas of Medicine were as closely defined as Dentistry and Veterinary sciences by Scopus, it is likely that most of them would aggregate close to Dentistry and Veterinary Science in the graph.

Other characteristics assessing the degree of isolation of researchers or research groups in different disciplines confirmed the gradients revealed in Figure 5. The citation impact (size of the bubbles) was directly proportional to the number of documents published per journal (Spearman Rank Correlation between Cit/Doc vs Doc/Jour: 0.66, p=0.002) ), and inversely proportional to the number of journals in each subject area (Spearman Correlation between Cit/Doc vs Journals: -0.54, p=0.02), if the data for the outlier data point Medicine was excluded. In addition, as shown in Figure 6, subject areas with journals with high number of publications, published papers with relatively lower country-self-citation rates (Spearman Correlation between (Pub/Jour vs CSC: 0.70, p=0.001). That is, subject areas with high average citation rates published more papers per journal, and those papers had relatively lower country-self-citations. Figure 6 shows a positive correlation between country self citation and citation impact (Spearman Correlation between CSC vs Cit/Doc: 0.51, p=0.03), with a similar gradient of academic areas as that revealed in Figure 5.

The same happened when we focus on International Collaboration (IC). Even if IC was heterogeneous between the geographical regions studied, the gradient of academic disciplines remained visible when comparing IC in the different regions (Figure 7). Humanities had the lowest IC everywhere, whereas Multidisciplinary Sciences and Physics in Western Europe, and Economics and Psychology in Asia, were the subject areas with the highest IC. International cooperation in Asia was lower compared to Western Europe for most areas except Arts & Humanities, Social Sciences, Psychology and Economics. Figure 7 shows only the 3 most prolific regions regarding scientific publications, but similar trends could be detected in the other regions. In 2011, the Pacific region, although among the lowest producers of scientific papers, had the highest IC and the Asian Region had the worst IC record.

**DISCUSSION**

Journals for multidisciplinary science, as computed by Scopus, publish selected research from mainly the natural sciences, such as Chemistry, Physics and Biology, that are deemed to be of interest to a broad range of scientists. Thus, multidisciplinary science, as defined here, is part of the natural sciences. On the other hand, psychology, business and economics are handled in most universities as part of the social sciences. The bibliographic data presented shows a bipolar gradient between these two groups of disciplines, where one extreme is represented by a cluster of areas in the natural sciences, and another extreme by a cluster of areas in the social sciences. The cluster that included natural sciences was constituted by publications that had high citation rates from colleagues publishing in relatively few journals containing a large number of articles. These publications cited relatively more research from countries other than the one from the author and had a higher proportion of international collaboration. In contrast, the cluster that included publications from the social sciences, had publications with relatively few citations and were published in many different journals that had a relatively low number of articles. These publications had relatively high country self citations and showed low levels of international collaboration.

These results can be explained in the light of the optimal strategies for swarm intelligence as revealed by the simulations. That is, the natural sciences conform more to exploratory strategies focusing on a few large knowledge clusters where "following" is more important than original new explorations. This leads to the existence few journals publishing many articles each. On the other hand, the social sciences conform more to an exploratory strategy optimizing search in many small isolated knowledge clusters, were "following" is less important than novel explorations in retrieving knowledge. This strategy seems to be have promoted the existence of many different journals in the social sciences, each with a relative small number of articles. In the natural sciences, trend following seems to be more rewarding than in the social sciences.

The simulation results suggest that a strategy that allocates research leaders rationally in accordance to the size of the knowledge clusters is optimal in all cases. Predicting the potential size of a knowledge cluster, however, is not easy and might be impossible in most cases. We do not know what remains to be discovered. Thus, in practice, we have to stick to less rational strategies. Policies favoring trend following over original researchers in the natural sciences and original independent researchers over trend followers in the social science would seem rational. The unconscious implementation of such policies seem to have occurred, possibly promoted by peer review, leading to different developments in different disciplines, explaining the actual bibliometric trends reported here.

It is curious to note that the 3 strategies explored in the simulations have been implemented by insect societies in their search for food [14]. Human scientists seem to have achieved the implementation of at least two of these strategies so far. Science policies, for example giving primary importance to citations [19] rather than to originality, might benefit certain areas or scientific communities more than others. More studies are needed to assess when and where this is desirable. But clearly, ways to quantify differences between the social and natural sciences are possible and should be developed further in order to gain a better understanding of their working dynamics.

The present study allowed us to get a glimpse of the knowledge landscape of different fields of knowledge. Analogous to explorations of fitness landscapes by genetic algorithms, different scientific disciplines explore different parts of our knowledge landscape and the adapted search strategy should reflect the structure of the landscape. Under this view, disciplines that have been qualified as more complex are also the ones with knowledge landscapes constituted by many small knowledge clusters. Humanities and social science, in this respect, seem to be more complex than chemistry and physics. Again, the index developed here could serve as a crude approximation to measuring these differences.

The main lesson from this exploration is that important differences in pursuing research exist and that interdisciplinary research has to understand these differences if it wants to expand successfully. The role of policymakers so far seems to be questionable. If scientific discovery is increasingly directed by policy makers (reducing our academic freedom to near zero), then we would not expect to find these differences between disciplines, or they should be converging, which they are not [20]. On the contrary, countries where policymakers avoid nudging the scientific activity in a specific direction seem to produce much better long term sustainable economic development than those favoring "strategic" areas [8]. Trail and error seem to have guided our scientific community relatively successful so far, differentiating the working of academic disciplines according to their tasks.

0066938

Data S1 http://atta.labb.usb.ve/Klaus/SupportingInformationPLOS2014.xls

**ACKNOWLEDGMENTS:** I thank Lui Lam, Xiau-Pu Han and Patrick Wessa for helpful comments on earlier drafts of the manuscript.

**FIGURE LEGENDS**

**Figure 1.** Amount of knowledge retrieved (K) with different proportions of original researchers (Ro) given as percentage of total researchers. Conditions were: TR system, D = 0.7 (points indicate the actual means from the 5 "runs" of the model; standard deviations were a maximum of 20 % of the means; curves were fitted by eye)

**Figure 2.** Relationship between the optimal number of Ro (ORo) as calculated with plots as shown in Figure 1, and the total amount of researchers available (N)

**Figure 3.** Relationship between the amount of knowledge retrieved (K) during 50 interactions of the simulation and the size of the knowledge clusters Kc for different recruitment techniques at D = 1

**Figure 4**. Amount of knowledge retrieved (K) by scientific communities using different recruitment techniques for Rt, for different values of Kc and D, so that the total knowledge in the system was constant: Kc x D = 50 (Kc was 1, 5, 15 and 500 respectively when D was 50, 10, 2 and 0.1)

**Figure 5:** Average number of papers per journal (Doc/Jour) plotted against the total number of journals registered by Scopus (Journals) for each of the subject area for the year 2011. The size of bubbles is proportional to the total number of citations for papers published 3 years earlier divided by the total number of papers published in that area as computed by SCImago for the year 2011.

**Figure 6**: Self-citation rates (CSC) for the year 2011 plotted against changes in citation impact of Cit/Doc during the same time period. The size of bubbles is proportional to Cit/Doc in 1999. The line shows the linear regression.

**Figure 7:** Level of international collaboration, measured as the proportion of document with affiliations from more than one country for Western Europe (WE), Asia and North America (proportional to the size of the bubble) during 2011.

**TABLES**

**Table 1**: Quantitative variables used

| IC | International Collaboration: Proportion of document with affiliations from more than one country |
|---|---|
| **Journals** | Number of Journals tracked by Scopus in a given subject category |
| **Doc/Jour** | Number of citable documents per journal in a given subject category |
| **Countries** | Number of countries reported in the addresses of the authors of the papers in that subject category |
| **Ref/Doc** | Number of references in the papers published in that subject category |
| **Cit/Doc** | Number of citations received during the following 3 years after publication by papers in that subject category |
| **CSC** | Level of provinciality, isolation or degree of country-self citation measured as the proportion of citations from the same country as the source paper. Country self-citations include author-self citations. |